# Disorder Dependent Valley Properties in Monolayer WSe$_2$

Kha Tran[1], Akshay Singh[1,*], Joe Seifert[1], Yiping Wang[1], Kai Hao[1], Jing-Kai Huang[2], Lain-Jong Li[2], Takashi Taniguchi[4], Kenji Watanabe[4], and Xiaoqin Li[1, 3+]

[1]Department of Physics and Center for Complex Quantum Systems, University of Texas at Austin, Austin, TX 78712, USA.
[2]Physical Science and Engineering Division, King Abdullah University of Science and Technology, Thuwal 23955, Saudi Arabia.
[3]Texas Materials Institute, University of Texas at Austin, Austin, TX 78712, USA.
[4]National Institute of Material Science, 1-1 Namiki, Tsukuba, Ibaraki 305-0044, Japan.
* Present address – Department of Material Science and Engineering, Massachusetts Institute of Technology, Cambridge, MA 02139, USA.

+E-mail address: elaineli@physics.utexas.edu

We investigate the effect of disorder on exciton valley polarization and valley coherence in monolayer WSe$_2$. By analyzing polarization properties of photoluminescence, the valley coherence (VC) and valley polarization (VP) is quantified across the inhomogeneously broadened exciton resonance. We find that disorder plays a critical role in the exciton VC, while affecting VP less. For different monolayer samples with disorder characterized by their Stokes Shift (SS), VC decreases in samples with higher SS while VP does not follow a simple trend. These two methods consistently demonstrate that VC as defined by the degree of linearly polarized photoluminescence is more sensitive to disorder, motivating further theoretical studies.

Valley refers to energy extrema in electronic band structures. Valley pseudo-spin in atomically thin semiconductors has been proposed and pursued as an alternative information carrier, analogous to charge and spin [1-7]. In monolayer transition metal dichalcogenides (TMDs), optical properties are dominated by excitons (bound electron-hole pairs) with exceptionally large binding energy and oscillator strength [8,9]. These excitons form at the energy extrema $K$ ($K'$) points at the Brillouin zone boundary. Due to broken inversion symmetry in combination with time-reversal symmetry, the valley and spin are inherently coupled in monolayer WSe$_2$. Valley contrasting optical selection rules make it possible to optically access and control the valley index via exciton resonances as demonstrated in valley specific dynamic Stark effect [10,11] as an example.

For valleytronic applications, particularly in the context of using valley as an information carrier, understanding both valley polarization and valley coherence are critical. Valley polarization represents the fidelity of writing information in valley index while valley coherence determines the ability to optically manipulate the valley index. Earlier experiments have demonstrated a high degree of valley polarization in photoluminescence (PL) experiments on some monolayer TMDs (e.g. MoS$_2$ and WSe$_2$), suggesting the valley polarization is maintained before excitons recombine [4-6,12]. Very recently, coherent nonlinear optical experiments have revealed a rapid loss of exciton valley coherence (~ 100 fs) in WSe$_2$ due to the intrinsic electron-hole exchange interaction [13]. The ultrafast dynamics associated with the valley depolarization (~ 1 ps) [14] and the even faster exciton recombination (~ 200 fs) [15,16] extracted from the nonlinear experiments are consistent with the PL experiments. As long as the valley depolarization and decoherence occurs on time scales longer or comparable with exciton recombination lifetime, steady-state PL signal shall preserve polarization properties reflecting the valley-specific excitations.

It is important to ask the question if disorder potential influences valley polarization and coherence, considering the fact that there are still significant amount of defects and impurities in these atomically thin materials and the substrate. This critical question has been largely overlooked in previous studies. Here, we investigate how valley polarization and coherence change in the presence of disorder potential. First, valley coherence is observed to change systematically across the inhomogeneously broadened exciton resonance while there is no clear trend in valley polarization.

Valley properties of an exfoliated monolayer WSe₂ encapsulated between hBN layers show a plateau beyond the mobility edge within the exciton resonance. We suggest that this systematic change is related to exciton localization by disorder potential, where the low energy side of the exciton resonance corresponds to weakly localized excitons and the high energy side is associated with more delocalized excitons [17,18]. Furthermore, we investigated a number of monolayer WSe₂ samples with different defect density characterized by the Stokes Shift (SS) between the exciton peak in photoluminescence and absorption. A higher degree of valley coherence is observed in samples with a smaller SS or lower defect density [19,20]. These two observations consistently suggest that shallow disorder potential reduces valley coherence. Our studies suggest that a more qualitative and systematic evaluation of valley coherence may guide the extensive on-going efforts in searching for materials with robust valley properties.

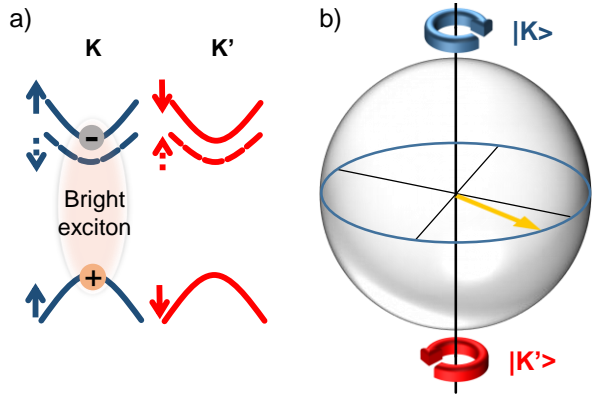

Figure 1 (color online): a) The band structure of monolayer WSe₂ at two degenerate K and K' valleys. A representative bright exciton transition is drawn at K valley between the highest valence band and second lowest conduction band. Valley contrasting spins allow left (right) circular polarized light to excite excitons in the K (K') valley. The next highest valence band with opposite spin is separated by ~300 meV and is omitted for clarity. b) Bloch Sphere representation of valley pseudospin degree of freedom. Circular polarized light prepares an exciton in |K> or |K'> state i.e. states at the poles, whereas linear polarized light prepares an exciton in a superposition of |K> and |K'> i.e. states at the equator.

The low energy bands with associated spin configurations in monolayer WSe₂ are illustrated in Fig. 1a. A dipole allowed (i.e., an optically bright) transition can only occur if the electron in the conduction and the missing electron in the valence band have parallel spins. Thus, the transition between the lowest conduction band and the highest valence band is dipole forbidden, and the lowest energy bright exciton transition is between the second lowest conduction band and the highest valence band as illustrated in Fig. 1a. Using $\sigma_+$ ($\sigma_-$) polarized excitation light, excitons are preferentially created in the $K$ ($K'$) valley due to the valley contrasting optical selection rules [1]. As with any binary quantum degree of freedom, K and K' valleys can be represented as a vector on a Bloch sphere, as shown in Fig. 1b. The degree of valley polarization is defined by the normalized difference in cross-circular and co-circular signals as

$$\rho_{VP} = (I_{co} - I_{cross})/(I_{co} + I_{cross}) \quad (1),$$

where $I_{co}(I_{cross})$ represents co (cross) circular polarized PL intensity with respect to the excitation polarization. Previous studies on monolayer WSe₂ have reported a large valley polarization in steady-state PL experiments [13,21] suggesting that the valley scattering rate is slower or comparable with exciton population recombination rate. In the Bloch sphere picture, a large VP suggests that once the Bloch vector is initialized along the north pole, it retains its orientation during exciton population recombination time. On the other hand, when a linearly polarized excitation laser is used, a coherent superposition of two valley excitons is created [21]. Such a coherent superposition state corresponds to a Bloch vector on the equatorial circle. Previous experiments suggest that exciton valley coherence, can be monitored by the linearly polarized PL signal [22,23]. Here, we follow this method and quantify the degree of valley coherence by the following definition

$$\rho_{VC} = (I_{col} - I_{cst})/(I_{col} + I_{cst}) \quad (2),$$

where $I_{col}(I_{cst})$ represents co (cross) linear polarized PL intensity with respect to the excitation polarization.

We first investigate the change of VC and VP as a function of energy across the exciton resonance on mechanically exfoliated monolayers WSe₂ sample E1 (Fig. 2c,d) and E5 (Fig. 2a,b). Sample E1 is a hBN encapsulated sample while sample E5 is a bare monolayer on sapphire substrate. The list of all samples studied and their labels are provided in the supplementary material [24]. It is known that the degree of valley polarization depends strongly on the excitation wavelength [21,25]. In our experiments, the excitation energy is chosen to be energetically close to the exciton resonance to observe a finite degree of VC, but far enough so that resonant Raman scattering does not interfere with VC [21,25]. Unless mentioned otherwise, for all PL measurements presented in this manuscript, we use a continuous wave laser at 1.88 eV (i.e. 660 nm) and keep the power ~ 20 µW at the sample with a focused spot size of ~ 2 µm diameter. All measurements were performed at temperature of ~ 13 K. A typical PL spectrum of monolayer WSe$_2$ exhibits two spectrally well resolved resonances corresponding to exciton and trion (a charged exciton), respectively (data included in supplementary [24]). There are two additional resonances at the lower energy, which may be due to either dark states or impurity bound states [26]. We focus on valley physics associated with the exciton resonance in this paper.

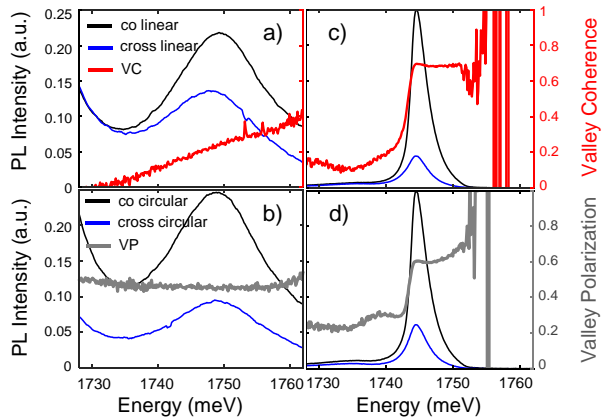

Figure 2 (color online): a) The exciton resonance shows co (cross) linear PL signal with respect to the linearly polarized excitation laser on an exfoliated WSe$_2$ monolayer. Corresponding VC is plotted on the right vertical axis. b) Co and cross circular polarized PL signal with respect to excitation polarization from the same sample in a). The VP is plotted on the right vertical axis. c) and d) Same measurements are repeated for a hBN encapsulated monolayer WSe$_2$ with a much sharper exciton linewidth ~ 4 meV.

Fig. 2a plots the co- and cross-linear polarized PL, and the corresponding VC across the exciton resonance. The VC drops to zero at the lower energy edge of the exciton resonance which spectrally overlaps with the trion resonance. Trions cannot exhibit VC in PL spectra due to photon-spin entanglement [21,27]. Interestingly, we observe a monotonic increase of the VC across the inhomogeneously broadened exciton resonance with $\rho_{VC}$ varying from 0 to 0.4 as shown in Fig. 2a. This monotonic change in VC across the exciton resonance is qualitatively repeated on all measured samples. VC reaches the maximum value at high energy side of the exciton and approaches zero at the low energy end. We suggest that the increase of VC across the exciton resonance arise from the degree of exciton localization [17,28,29].

In contrast, VP remains constant across the exciton resonance with $\rho_{VP}$ ~ 0.48 as illustrated in Fig. 2b. Previous studies suggested that only atomically sharp potentials can induce inter-valley scattering and depolarization of valley exciton [21]. Thus, the nearly constant VP suggests that the inhomogeneously broadened exciton resonance is mainly due to slowly varying spatial potentials (in contrast to atomically sharp potentials). Such disorder potential may be attributed to local strain as well as shallow impurity potentials [17,28,29]. This speculation is also consistent with the observation that strongly localized excitons likely due to deep, atomically sharp potentials appear at much lower energy, ~ 100-200 meV below the exciton resonance [30,31]. An important mechanism causing valley depolarization is electron-hole exchange unaffected by shallow potential fluctuations [32-34]. Other valley scattering mechanisms such as Dyakanov-Perel (DP) and Eliott-Yafet (EY) mechanisms are slower and considered unimportant for excitons in TMDs [32].

Next, we study the VP and VC of a hBN encapsulated monolayer WSe$_2$ with a narrow exciton PL linewidth of ~ 4 meV. Fig. 2c-d plots the degree of valley coherence and valley polarization across the exciton resonance. We observe a very sharp rise of VC at low energy side of the exciton followed by a plateau of ~ 0.7 near the exciton peak and beyond. The VC plateau suggests that a mobility edge exists within the exciton resonance in this high quality sample. At energy below the mobility edge, weakly localized excitons experience gradually varying disorder potential. Beyond the mobility edge, mostly delocalized excitons exhibit the same degree of VC. The qualitative feature of VP across the exciton resonance follows a similar trend, a rapid rise transitioning to a plateau of ~0.6 near the exciton peak. The rise of the VP in this sample is not consistent with constant VP observed in other WSe$_2$ monolayer samples. It may arise from other low energy resonances that are spectrally distinct in the encapsulated sample with a narrow exciton resonance.

To further investigate the role of disorder on valley properties, we studied a total of eight monolayer WSe$_2$ samples. We assume that the defect density is correlated with the spectral shift between exciton resonances measured in PL and absorption, known as the Stokes Shift (SS). As a simple method based entirely on commonly used linear optical spectroscopy methods, SS has been used to characterize a wide variety of systems [19,35] including defect density [36-38] and thickness fluctuations in quantum wells [20,39,40] and size distribution in ensembles of quantum dots [41,42].

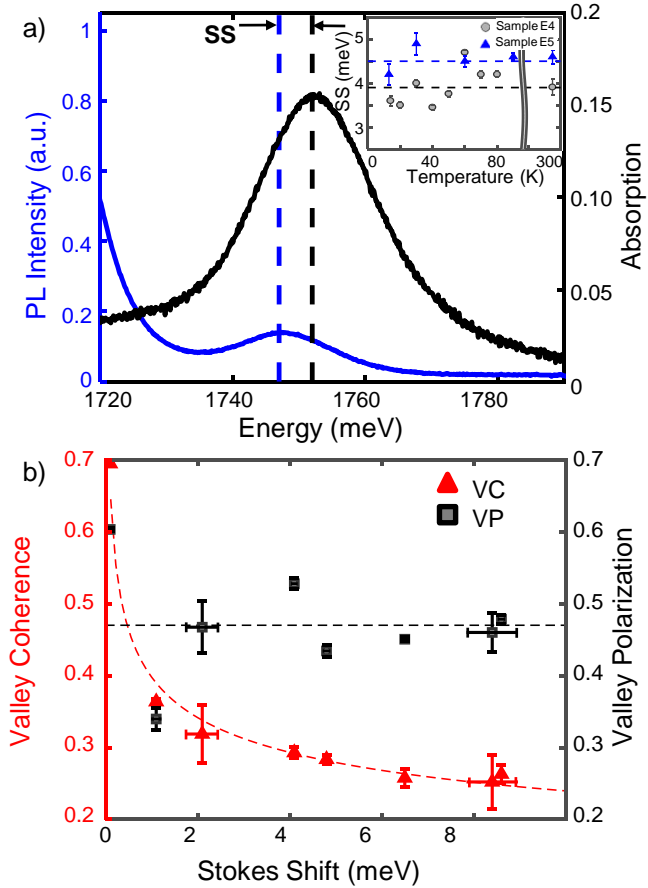

Figure 3 (color online): a) Stokes shift is shown as the difference in energy between the absorption spectrum and PL from the exciton resonance. Inset: SS dependence on temperature b) VC (VP) is plotted with respect to SS. VC shows an inverse dependence versus SS whereas VP shows no recognizable trend.

A typical SS measurement is shown in Fig. 3a. The PL and white light absorption spectra are taken from the same exfoliated monolayer WSe$_2$ labeled E5. The absorption spectrum is plotted as a differential and normalized spectrum $(T_s - T_m)/T_s$, where $T_s$ is the transmission through the substrate and $T_m$ is the transmission through both the substrate and monolayer sample. The exciton resonances in the PL and absorption are fitted with Gaussian functions. The peaks extracted from the fittings are indicated by the dotted lines, yielding a 4.8 meV SS for this sample. As we increase the temperature from 13K to room temperature, SS varies within ±0.6 meV (inset of Fig. 3a), which is within the error bars of our measurements.

To quantify the dependence of valley properties on SS (and on disorder potentials), the above measurements are repeated on all eight samples. For comparison across different samples, the VC (or VP) value for each sample is calculated by taking the average of the VC (or VP) in a range spanning ±$\sigma/6$ from the exciton peak where $\sigma$ is the fitted linewidth. We found the range of the spectral integration does not change our qualitative conclusion. Vertical error bars for VC (VP) in Fig. 3b are the standard deviation of the VC (VP) values. Horizontal error bars for SS in Fig. 3b are the sum of the fitting errors of the absorption spectrum and the PL spectrum. Some samples have small fitting errors and therefore their error bars are not visible within the plotted SS range. The results as summarized in Fig. 3b have a number of interesting features. Firstly, VC is found to decrease significantly with increasing SS of samples, with a fractional drop of ∼ 64 % between the samples with the lowest to highest SS. Specifically, $\rho_{VC}$ varies from 0.69 to 0.25 as SS changes from 0 meV to 8.6 meV. Secondly, $\rho_{VP}$ varies from 0.43 to 0.6 across the samples, and no clear correlation between VP and SS is observed. Based on the assumption that SS is correlated with the defect density in different samples, we infer that disorder potential reduces VC but has no clearly identified systematic influence on VP. This conclusion is consistent with the spectral dependence of VC and VP across the exciton resonance observed on individual samples as reported in Fig. 2a-d. In addition, a recent experiment [28] investigated spatial variations of VP and VC on a CVD grown monolayer WSe$_2$. While VP was found to be mostly constant, VC showed significant changes, likely arising from disorder potential.

In summary, we report an experimental study of the effect of disorder on VC and VP in monolayer WSe$_2$. The low energy side of the exciton resonance is associated with weakly localized excitons, and the high energy side with more delocalized excitons. Using steady state polarization resolved PL, we observe that the VC increases across the inhomogeneously broadened exciton resonance. In the highest quality exfoliated sample encapsulated between hBN layers, a plateau of VC is reached at energy beyond a mobility edge within the exciton resonance. The existence of a mobility edge within exciton resonance and its effect on trion formation dynamics have been previously studied in monolayer MoSe$_2$ [17]. VP and VC are then measured for a number of samples with different SS (a measure of disorder). VC varies inversely with SS while no clear and systematic changes of VP have been observed across different samples. Our observations suggest that shallow disorder potentials have a crucial effect on the exciton valley coherence. Particularly, weakly localized excitons lose valley coherence more rapidly than the delocalized excitons. A recent theoretical study suggested that the exciton intervalley scattering time from disorder potential does not exhibit monotonic changes [43]. However, scattering time cannot be directly extracted from steady-state PL experiments reported here. Our work should motivate future experiments and microscopic theoretical studies necessary for a comprehensive understanding of the effect of disorder on valley properties in TMDs.

**Acknowledgements**: We gratefully acknowledge helpful discussions with Galan Moody and Fengcheng Wu. The spectroscopic experiments were jointly supported by NSF DMR-1306878 (A. Singh) and NSF EFMA-1542747 (K. Tran, K. Hao, J. Seifert, and X. Li). Li also gratefully acknowledge support from Welch Foundation grant F-1662. K.W. and T.T. acknowledge support from the Elemental Strategy Initiative conducted by the MEXT, Japan and JSPS KAKENHI Grant Number JP15K21722.

## References

[1]    D. Xiao, G.-B. Liu, W. Feng, X. Xu, and W. Yao, Phys. Rev. Lett. **108**, 196802 (2012).

[2]    J. Lee, K. F. Mak, and J. Shan, Nat. Nano. **11**, 421 (2016).

[3]    K. F. Mak, K. L. McGill, J. Park, and P. L. McEuen, Science **344**, 1489 (2014).

[4]    H. Zeng, J. Dai, W. Yao, D. Xiao, and X. Cui, Nat Nano **7**, 490 (2012).

[5]    T. Cao, G. Wang, W. Han, H. Ye, C. Zhu, J. Shi, Q. Niu, P. Tan, E. Wang, B. Liu, and J. Feng, Nat Comm **3**, 887 (2012).

[6]    G. Sallen, L. Bouet, X. Marie, G. Wang, C. R. Zhu, W. P. Han, Y. Lu, P. H. Tan, T. Amand, B. L. Liu, and B. Urbaszek, Phys. Rev. B **86**, 081301 (2012).

[7]    R. Schmidt, A. Arora, G. Plechinger, P. Nagler, A. Granados del Águila, M. V. Ballottin, P. C. M. Christianen, S. Michaelis de Vasconcellos, C. Schüller, T. Korn, and R. Bratschitsch, Phys. Rev. Lett. **117**, 077402 (2016).

[8]    K. F. Mak, C. Lee, J. Hone, J. Shan, and T. F. Heinz, Phys. Rev. Lett. **105**, 136805 (2010).

[9]    A. Splendiani, L. Sun, Y. Zhang, T. Li, J. Kim, C. Y. Chim, G. Galli, and F. Wang, Nano Letters **10**, 1271 (2010).

[10]   E. J. Sie, J. W. McIver, Y.-H. Lee, L. Fu, J. Kong, and N. Gedik, Nat Mater **14**, 290 (2015).

[11]   J. Kim, X. Hong, C. Jin, S.-F. Shi, C.-Y. S. Chang, M.-H. Chiu, L.-J. Li, and F. Wang, Science **346**, 1205 (2014).

[12]   K. F. Mak, K. He, J. Shan, and T. F. Heinz, Nat Nano **7**, 494 (2012).

[13]   K. Hao, G. Moody, F. Wu, C. K. Dass, L. Xu, C.-H. Chen, L. Sun, M.-Y. Li, L.-J. Li, A. H. MacDonald, and X. Li, Nat Phys **12**, 677 (2016).

[14]   A. Singh, K. Tran, M. Kolarczik, J. Seifert, Y. Wang, K. Hao, D. Pleskot, N. M. Gabor, S. Helmrich, N. Owschimikow, U. Woggon, and X. Li, Phys. Rev. Lett. **117**, 257402 (2016).

[15]   C. Poellmann, P. Steinleitner, U. Leierseder, P. Nagler, G. Plechinger, M. Porer, R. Bratschitsch, C. Schuller, T. Korn, and R. Huber, Nat Mater **14**, 889 (2015).

[16]   G. Moody, C. Kavir Dass, K. Hao, C.-H. Chen, L.-J. Li, A. Singh, K. Tran, G. Clark, X. Xu, G. Berghäuser, E. Malic, A. Knorr, and X. Li, Nat Comm **6**, 8315 (2015).

[17]   A. Singh, G. Moody, K. Tran, M. E. Scott, V. Overbeck, G. Berghäuser, J. Schaibley, E. J. Seifert, D. Pleskot, N. M. Gabor, J. Yan, D. G. Mandrus, M. Richter, E. Malic, X. Xu, and X. Li, Phys. Rev. B **93**, 041401(R) (2016).

[18]   A. Hichri, B. Amara, A. Sabrine, and J. Sihem, arXiv:1609.05634  (2016).

[19]   F. Yang, M. Wilkinson, E. J. Austin, and K. P. O'Donnell, Phys. Rev. Lett. **70**, 323 (1993).

[20]   F. Yang, P. J. Parbrook, B. Henderson, K. P. O'Donnell, P. J. Wright, and B. Cockayne, Appl. Phys. Lett. **59**, 2142 (1991).

[21]   A. M. Jones, H. Yu, N. J. Ghimire, S. Wu, G. Aivazian, J. S. Ross, B. Zhao, J. Yan, D. G. Mandrus, D. Xiao, W. Yao, and X. Xu, Nat Nano **8**, 634 (2013).

[22]   Z. Ye, D. Sun, and T. F. Heinz, Nat Phys **13**, 26 (2017).

[23]   G. Wang, X. Marie, B. L. Liu, T. Amand, C. Robert, F. Cadiz, P. Renucci, and B. Urbaszek, Phys. Rev. Lett. **117**, 187401 (2016).

[24]   See Supplementary Material for details.

[25]   G. Wang, M. M. Glazov, C. Robert, T. Amand, X. Marie, and B. Urbaszek, Phys. Rev. Lett. **115**, 117401 (2015).

[26]   H. Dery and Y. Song, Phys. Rev. B **92**, 125431 (2015).

[27]   K. Hao, L. Xu, F. Wu, P. Nagler, K. Tran, X. Ma, C. Schuller, T. Korn, A. H. MacDonald, G. Moody, and X. Li, arXiv:1611.03388  (2016).

[28]   A. Neumann, J. Lindlau, L. Colombier, M. Nutz, S. Najmaei, J. Lou, A. D. Mohite, H. Yamaguchi, and A. Högele, Nat Nano **DOI: 10.1038/nnano.2016.282** (2017).

[29]   T. Jakubczyk, V. Delmonte, M. Koperski, K. Nogajewski, C. Faugeras, W. Langbein, M. Potemski, and J. Kasprzak, Nano Letters **16**, 5333 (2016).

[30]     A. Srivastava, M. Sidler, A. V. Allain, D. S. Lembke, A. Kis, and A. Imamoğlu, Nat Nano **10**, 491 (2015).

[31]     Y.-M. He, G. Clark, J. R. Schaibley, Y. He, M.-C. Chen, Y.-J. Wei, X. Ding, Q. Zhang, W. Yao, X. Xu, C.-Y. Lu, and J.-W. Pan, Nat Nano **10**, 497 (2015).

[32]     T. Yu and M. W. Wu, Phys. Rev. B **89**, 205303 (2014).

[33]     M. Z. Maialle, E. A. de Andrada e Silva, and L. J. Sham, Phys. Rev. B **47**, 15776 (1993).

[34]     A. Ramasubramaniam, Phys. Rev. B **86**, 115409 (2012).

[35]     X. Qian, Y. Zhang, K. Chen, Z. Tao, and Y. Shen, Dyes and Pigments **32**, 229 (1996).

[36]     S. Chichibu, J. Vac. Sci. Technol. B **16**, 2204 (1998).

[37]     P. R. Kent and A. Zunger, Phys. Rev. Lett. **86**, 2613 (2001).

[38]     S. Srinivasan, F. Bertram, A. Bell, F. A. Ponce, S. Tanaka, H. Omiya, and Y. Nakagawa, Appl. Phys. Lett. **80**, 550 (2002).

[39]     L. C. Andreani, G. Panzarini, A. V. Kavokin, and M. R. Vladimirova, Phys. Rev. B **57**, 4670 (1998).

[40]     O. Rubel, M. Galluppi, S. D. Baranovskii, K. Volz, L. Geelhaar, H. Riechert, P. Thomas, and W. Stolz, J. Appl. Phys. **98**, 063518 (2005).

[41]     B. L. Wehrenberg, C. Wang, and P. Guyot-Sionnest, J. Phys. Chem. B **106**, 10634 (2002).

[42]     A. Franceschetti and S. T. Pantelides, Phys. Rev. B **68**, 033313 (2003).

[43]     T. Yu and M. W. Wu, Phys. Rev. B **93**, 045414 (2016).

# Supplementary for Disorder Dependent Valley Properties in Monolayer WSe₂

Kha Tran[1], Akshay Singh[1,*], Joe Seifert[1], Yiping Wang[1], Kai Hao[1], Jing-Kai Huang[2], Lain-Jong Li[2], Takashi Taniguchi[4], Kenji Watanabe[4], and Xiaoqin Li[1, 3+]

[1]Department of Physics and Center for Complex Quantum Systems, University of Texas at Austin, Austin, TX 78712, USA.
[2]Physical Science and Engineering Division, King Abdullah University of Science and Technology, Thuwal 23955, Saudi Arabia.
[3]Texas Materials Institute, University of Texas at Austin, Austin, TX 78712, USA.
[4]National Institute of Material Science, 1-1 Namiki, Tsukuba, Ibaraki 305-0044, Japan.
* Present address – Department of Material Science and Engineering, Massachusetts Institute of Technology, Cambridge, MA 02139, USA.

## I. Experimental Setup

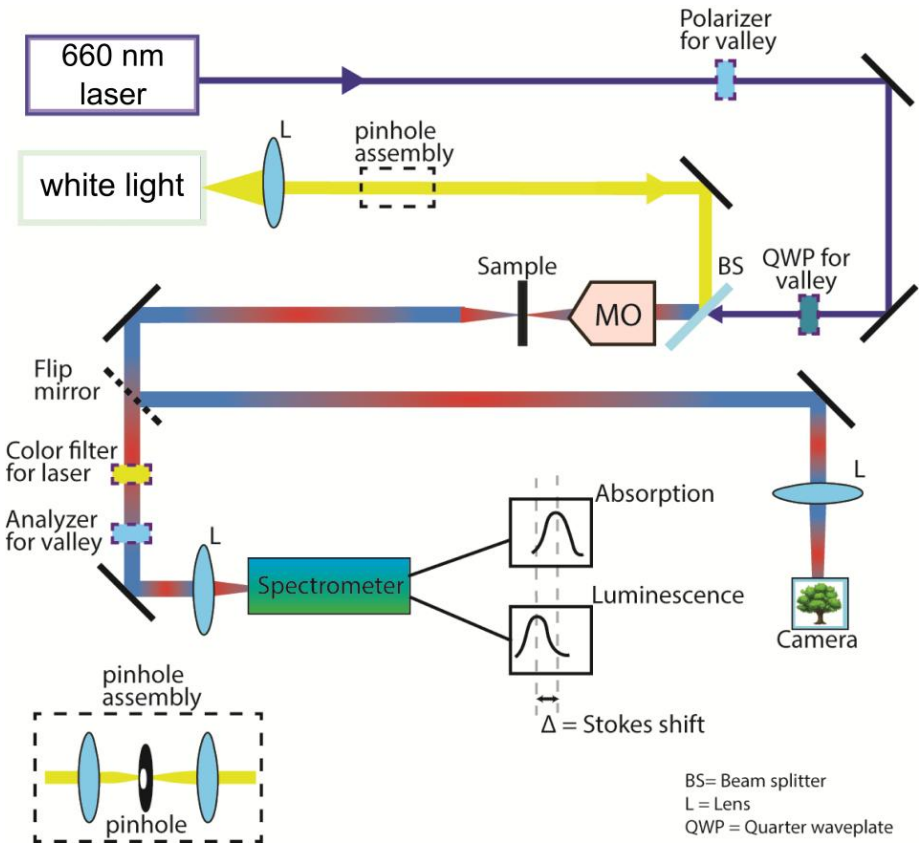

Figure S1: Optical setup for measuring SS and valley properties of monolayer WSe2

The optical setup schematic is depicted in figure S1. 660 nm continuous wave laser is focused onto the sample using a 50X Mitutoyo objective whose numerical aperture is 0.42. The laser spot diameter at the sample is 2 µm. To achieve a similar spot size with the white light, we use a pair of 5 cm focusing lenses together with a 100 µm pinhole inserted at the focal plane in between the lenses (inset Fig. S1). To measure Stokes Shift, the PL excited with red laser and the white light absorption spectrum are collected via transmission geometry. To measure valley properties, the polarized resolved PL spectra are collected via reflection geometry.

## II. Fitting Comparison of Absorption Spectrum and Sample Information

We have investigated 8 samples. They are labeled E1, E2, CVD3, E4, E5, E6, E7, E8, and CVD9.  We use one CVD sample and take data at two different sample locations: hence CVD3 and CVD9. The PL spectra collected on the CVD sample showed large variations at different locations and the PL intensity is also significantly weaker than that from exfoliated samples. Although we have included data from one particular location on the CVD sample with a small SS, it should not be interpreted as a better quality sample  compared to other exfoliated samples. CVD9 data is omitted in the main text but is included the supplementary section V for completeness. Sample E1 is the encapsulated monolayer with 4 meV PL linewidth and zero SS. All other samples: E2 to E8 are bare monolayers on sapphire substrate. The sample order is arranged in table S1 so that they are in order of increasing Stoke Shift.

We have fit absorption profiles with three different lineshapes: gaussian, lorentzian and half gaussian. In half gaussian fitting method, only data points from the FWHM to the peak are fitted with Gaussian. The comparison of the three methods is summarized below in Table S1. In Fig. S2, we also show an example of the absorption lineshape fitted with the three methods. The SS measured with three method showed smaller variations than the error bars quoted in main text. To be consistent, we always quoted the SS calculated from the Gaussian fitting method.

| Sample | Peak position (meV) | | | FWHM (meV) | | | Stokes Shift (SS) | | |
|---|---|---|---|---|---|---|---|---|---|
|  | *L* | *G* | *Half-G* | *L* | *G* | *Half-G* | *L* | *G* | *Half-G* |
| **E1** | 1744.5 | 1744.7 | 1744.5 | 9.1 | 8.26 | 8.98 | -0.1 | 0.1 | -0.1 |
| **E2** | 1745.3 | 1745.4 | 1745.4 | 14.5 | 11.03 | 11.03 | 1 | 1.1 | 1.1 |
| **CVD3** | 1743.5 | 1744 | 1743.7 | 23.1 | 20.7 | 23.7 | 1.6 | 2.1 | 1.8 |
| **E4** | 1755.8 | 1755.8 | 1755.7 | 17.6 | 14.9 | 13.6 | 4.1 | 4.1 | 4.0 |
| **E5** | 1757.2 | 1757.3 | 1757.2 | 18.1 | 15.9 | 12.8 | 4.7 | 4.8 | 4.7 |
| **E6** | 1753.7 | 1753.7 | 1753.6 | 20.8 | 16.1 | 15.4 | 6.5 | 6.5 | 6.5 |
| **E7** | 1755.7 | 1756.6 | 1756.6 | 44.7 | 36.8 | 25.0 | 7.5 | 8.4 | 8.3 |
| **E8** | 1757.5 | 1757.5 | 1757.1 | 21.1 | 17.0 | 15.5 | 8.6 | 8.6 | 8.3 |
| **CVD9 (omitted)** | 1713.7 | 1715.2 | 1714.7 | 48.7 | 45.4 | 48.0 | 26.0 | 27.5 | 27 |

Table S1: Summary of absorption spectrum peak positions, full-width half-maximum (FWHM) and Stokes Shift (SS), using Lorentzian (L), Gaussian (G) and Half-gaussian (Half-G) fitting methods, for different samples.

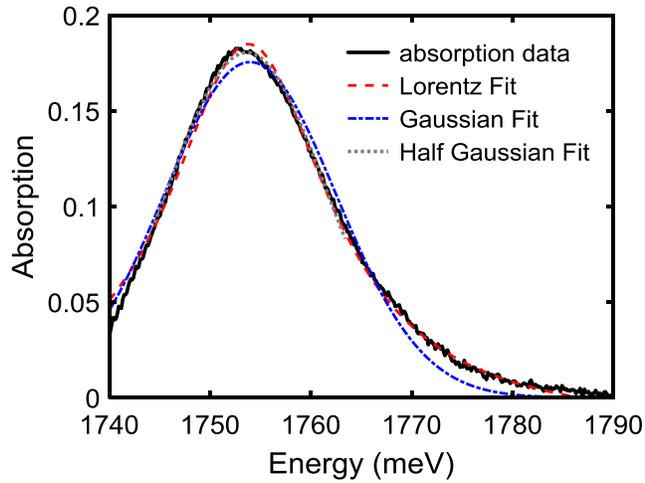

Figure S2: Absorption linewidth of sample E6 fitted using different fitting methods: Lorentz, Gauss, and half Gauss.

### III. Stokes Shift Plotted Against Absorption Linewidth

We fit Gaussian profiles to exciton absorption spectra and plot SS versus FWHM of the fitted Gaussian for all 8 monolayer samples. The vertical error bars of SS are due to the combined fitting errors of both PL and absorption peak. The horizontal error bars of FWHM are small and therefore not visible on the scale plotted. The correlation between SS and FWHM is only valid over a certain range of the FWHM. We speculate that the lack of correlation between the two quantities could be due to different types of defects causing inhomogeneous broadening in different samples.

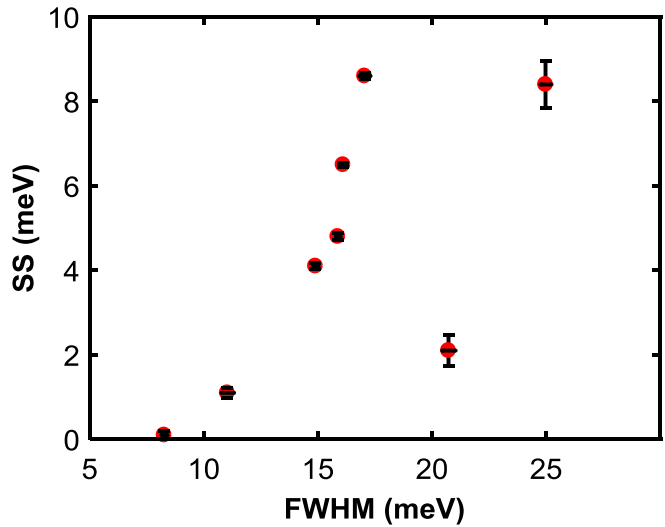

Figure S3: The absorption spectra are fitted with Gaussian profile and the FWHMs are extracted and plotted against SS.

## IV. Subtraction of Trion Contribution from Exciton Degree of Valley Coherence

The data shown in Fig. S4 and data in the main text Fig. 2a-b are from the same exfoliated sample E5 whose SS is 4.8 meV. Here, we plot the data over a greater energy range to show the trion resonances. We fit the trion resonances of co and cross linear PL signals with gaussians. We then subtract the trion fitting curve from co and cross PL signals and compute the degree of valley coherence from exciton. Evidently, the degree of VC computed before and after the trion subtraction is the same.

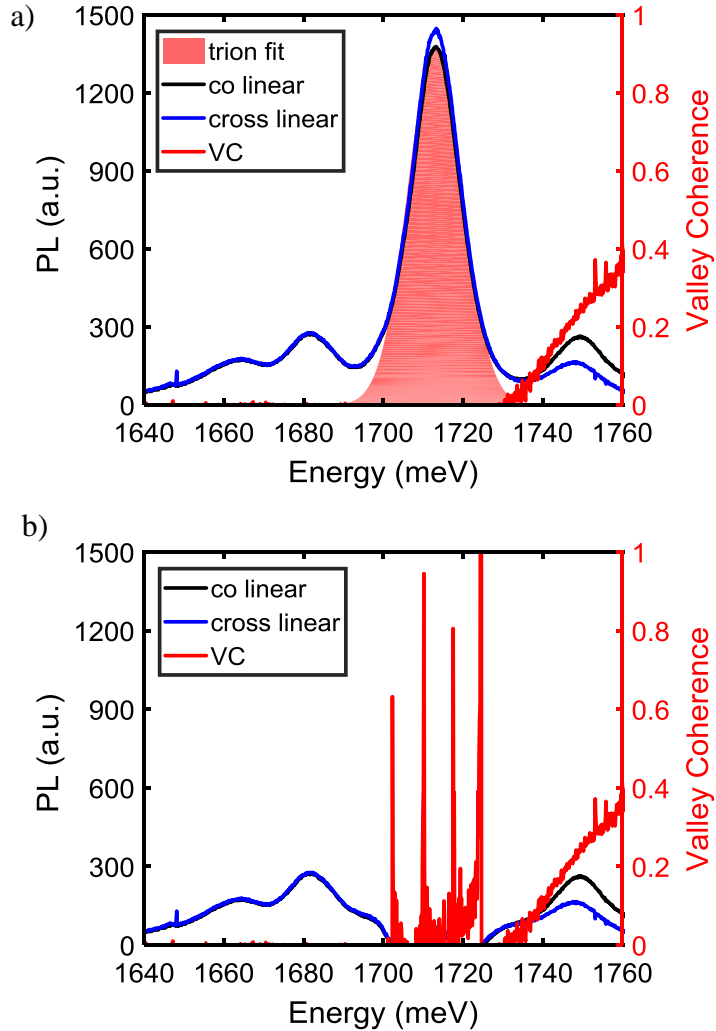

Figure S4: a) Trion resonances on both co and cross linear signal are fitted with Gaussians. Since trion VC cannot be detected in PL experiments, the fits from co and cross linear polarized PL spectra are essentially the same. VC is shown here before the trion subtraction from the co and cross linear PL signals. b) After trion subtraction, the valley coherence is unchanged, signifying that the trion resonance has minimal contribution to the analysis of exciton VC.

## V. Discussion of an Additional Data Point from CVD Sample

We have omitted a data point with large SS of 27.5 meV taken on the CVD sample whose exciton resonance peak is at 1690 meV. This particularly large SS likely originated from a location on the sample with particularly large defect density. However, this low resonance peak might also be influenced by others states such as trions or impurity states. Thus, we omitted this data point in the main text.

For completeness, we show data for this particular location on the CVD sample with a particularly large SS of 27.5 meV and low PL exciton energy peak ~1690 meV. The PL linewidth (full width half max) is 39.6 meV. However, the VC still shows the characteristic increase with energy and VP is relatively flat around the exciton peak, consistent with the qualitative features reported in Fig. 2a-d. Fig. S4c shows the variation of VC and VP versus SS including the omitted data point. The trends are fully consistent with the discussion in the main text.

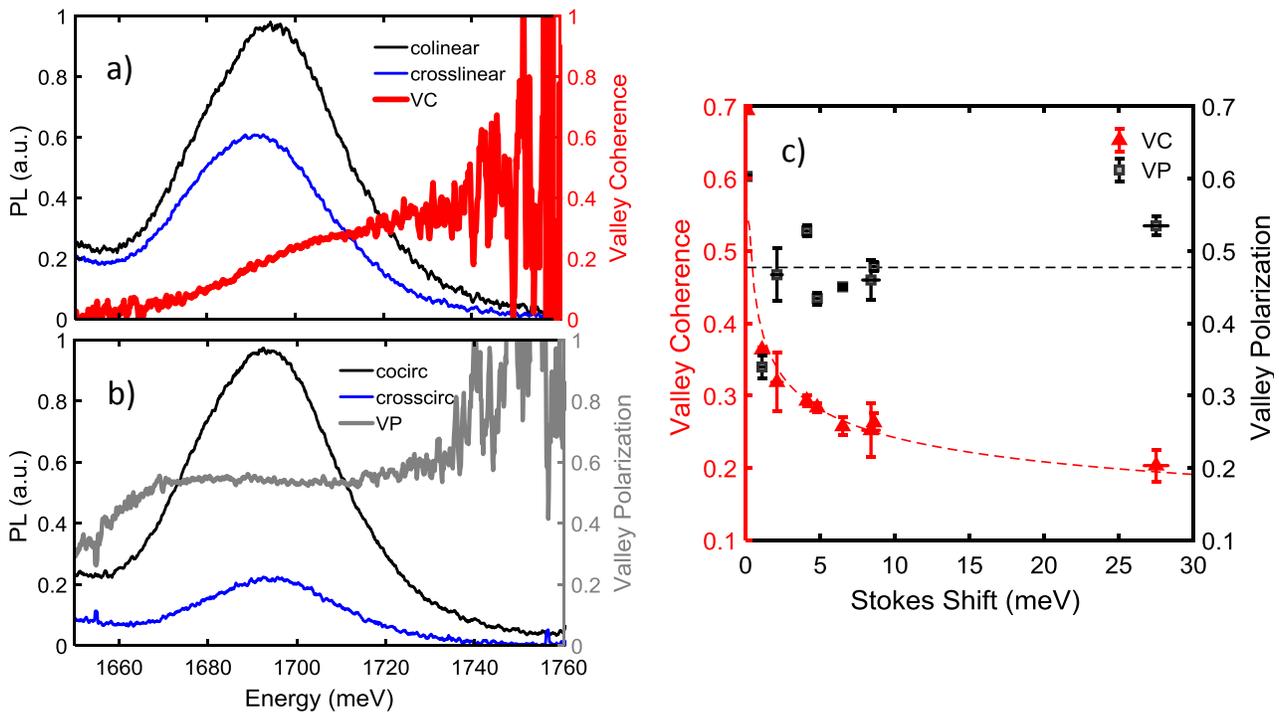

Figure S5: Degree of a) valley coherence and b) valley polarization plotted across the exciton resonance of omitted CVD sample. c) Valley coherence and valley polarization plotted against SS included the omitted data point.